\documentclass[technote,10pt]{IEEEtran}
\usepackage{flushend}
\usepackage{amsmath,amsfonts}
\usepackage{algorithm}
\usepackage{algpseudocode}
\usepackage{color}
\usepackage{booktabs}
\usepackage{array}
\usepackage[caption=false,font=normalsize,labelfont=sf,textfont=sf]{subfig}
\usepackage{textcomp}
\usepackage{stfloats}
\usepackage{cite, url}
\usepackage{verbatim}
\usepackage{graphicx}
\usepackage{hyperref}
\usepackage{bm}
\begin{document}

\title{
Generative Video Semantic Communication via Multimodal Semantic Fusion with Large Model
}
\author{
Hang Yin, Li Qiao, Yu Ma, Shuo Sun, Kan Li$^{*}$, Zhen Gao$^{*}$, and Dusit Niyato~\IEEEmembership{Fellow,~IEEE}
\thanks{The work was supported by the Natural Science Foundation of China (NSFC) under Grant 62471036 and U223320067, Shandong Province Natural Science Foundation under Grant ZR2022YQ62, Beijing Natural Science Foundation under Grant L242011, and Grant YQ24167, and Beijing Nova Program. (\textit{$^{*}$Corresponding authors: Zhen Gao, Kan Li.)}}
\thanks{Hang Yin, Li Qiao, Yu Ma, Shuo Sun, Kan Li$^{*}$, and Zhen Gao$^{*}$ are with the School of Information and Electronics, Beijing Institute of Technology, Beijing 100081, China (e-mail: \{yinhang, qiaoli, yu.ma, sunshuo2002, likan, gaozhen16\}@bit.edu.cn).}
\thanks{Dusit Niyato is with the School of Computer Science and Engineering, Nanyang Technological University, Singapore 639798 (e-mail: dniyato@ntu.edu.sg).}
}

\maketitle
\begin{abstract}
Despite significant advancements in traditional syntactic communications based on Shannon's theory, these methods struggle to meet the requirements of 6G immersive communications, especially under challenging transmission conditions. With the development of generative artificial intelligence (GenAI), progress has been made in reconstructing videos using high-level semantic information. In this paper, we propose a scalable generative video semantic communication framework that extracts and transmits semantic information to achieve high-quality video reconstruction. Specifically, at the transmitter, {\color{black}description and other condition signals (e.g., first frame, sketches, etc.) are extracted from the source video, functioning as text and structural semantics, respectively.} At the receiver, the {\color{black}diffusion-based GenAI large models} are utilized to fuse the semantics of the {\color{black}multiple} modalities for reconstructing the video. Simulation results demonstrate that, at an ultra-low channel bandwidth ratio (CBR), our scheme effectively captures semantic information to reconstruct videos aligned with human perception under different signal-to-noise ratios. {\color{black}Notably, the proposed \textbf{First Frame+Desc.} scheme consistently achieves CLIP score exceeding 0.92 at CBR = 0.0031 for SNR $>$ 0\,dB. This demonstrates its robust performance even under low SNR conditions.}
\end{abstract}

\section{Introduction}
Semantic communication is considered a revolutionary paradigm with the potential to transform the design and operation of 6G wireless communication systems~\cite{pokhrel2022understand,wang2022transformer, liu2024near,wu2024deep}.
Whereas, extracting semantics from source signals and redesigning wireless communication systems present significant challenges. Facing such challenges, the advanced coding techniques, such as deep joint source-channel coding (DJSCC), have been proposed and achieved success in semantic communication systems, where the semantic compression capability of neural networks is leveraged in an end-to-end training manner~\cite{DBLP:journals/tccn/BourtsoulatzeKG19, DBLP:journals/jsac/TungG22}. However, the DJSCC frameworks rely on rate-distortion theory and are unable to optimize the perception quality, which is crucial to humans.
 
On the other hand, advancements in generative artificial intelligence (GenAI) not only enable the creation of realistic, high-quality content but also facilitate its efficient transmission through semantic communication innovations, thereby optimizing bandwidth usage without compromising on quality.
Early GenAI focused on probabilistic graphical models, such as hidden Markov models~\cite{juang1986maximum}, which generated data with limited quality and diversity. Variational autoencoders (VAEs) marked a breakthrough by producing more complex and realistic data~\cite{kingma2013auto,dang2019real}, but they faced distribution mismatch issues. To address this, generative adversarial networks (GANs) were proposed~\cite{yu2017seqgan,zhang2017adversarial}, optimizing the generator and discriminator through adversarial loss to improve data quality. GANs have been used after DJSCC at the receiver to enhance perception quality~\cite{erdemir2023generative}. However, GANs are prone to instability, such as mode collapse, resulting in a lack of diversity in the generated samples.

Recently, many large models using diffusion techniques have appeared, and they can generate data based on conditional information.~\cite{DBLP:conf/nips/SahariaCSLWDGLA22,brooks2024video}. 
For example, the prompt \textit{``Panda plays ukelele at home''} can be used to generate a short video whose semantic content aligns with the description provided in the prompt~\cite{DBLP:journals/corr/abs-2401-12945}. Due to the generative models' advantages in accurate grasp of semantic conditions and the substantial improvement of content authenticity, the authors in~\cite{DBLP:journals/corr/abs-2405-09976} and~\cite{qiao2024latency} focused on extracting and transmitting compressed semantic features, such as text and edge maps, to guide the diffusion process at the receiver. They demonstrated that even at ultra-low compression rates (less than 0.1 bit per pixel), essential semantics can still be preserved at the receiver. However, these works focus on image transmission, and the research on video transmission is still lacking. The authors in~\cite{DBLP:journals/corr/abs-2402-08934} and~\cite{liu20242} explored the use of diffusion models to reconstruct inter-frame information in video frames, focusing primarily on video compression in error-free channels. 

Therefore, a significant research gap exists in effectively utilizing diffusion models for video reconstruction under practical wireless transmission channels. 
Delivering video semantics as conditional signals for diffusion models to achieve high semantic scores remains an open challenge. 

To fill these gaps, we design a generative video semantic communication (GVSC) framework for transmitting video semantics and reconstructing videos, whereby GenAI large models with the diffusion techniques are utilized to generate video aligned with human perception from video semantics of sketch sequences and text description. 
Our contributions are summarized as follows:

1) We propose a novel GVSC framework for video reconstruction at ultra-low channel bandwidth ratios (CBR). {\color{black}At the transmitter, our semantic extractor identifies key visual and textual semantics from the video. The visual information is transmitted via DJSCC, while the textual information uses turbo coding with quadrature amplitude modulation (QAM).} These schemes include critical details such as object location, color, size, and actions. Unlike traditional schemes, GVSC leverages a pre-trained GenAI model to reconstruct video content from condensed semantic data. At the receiver, a GenAI diffusion model integrates this information for effective video reconstruction. The robustness of GVSC ensures effectiveness even in resource-constrained environments, performing well at CBR as low as $10^{-2}$.

2) We introduce several video semantic extraction strategies that adapt the transmission schemes according to signal-to-noise ratio (SNR) conditions. 
{\color{black}Our approach includes a variety of strategies, such as single-sketch with video description, sketch sequence combined with video descriptions, and incorporating the first red-green-blue (RGB) frame with video descriptions. By systematically evaluating these strategies under varying SNR scenarios, our generative transmission schemes effectively mitigate the cliff effect~\cite{proakis2008digital}. Moreover, these models exhibit strong adaptability in the low SNR range, ensuring reliable video communication even under challenging conditions.}

3) We design a weighted loss function that combines mean squared error loss (MSE) and learned perceptual image patch similarity (LPIPS) loss to optimize the transmission quality of sketch sequence semantics. Our simulations demonstrate that using the proposed weighted loss function aligns more closely with human perception.

\begin{figure*}[ht] 
\centering 
\includegraphics[trim=290 240 245 235, clip,scale=1.6]{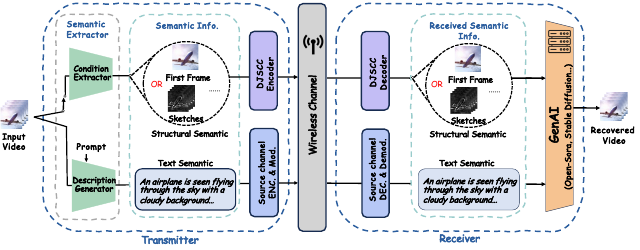} 

\caption{{\color{black}\textbf{GVSC framework overview.} We transmit key semantic information, including visual and textual elements, to capture the main semantic content of the video. This semantic data is then fused by the GenAI large model at the receiver to accurately reconstruct the video.}
}
\label{fig:GVSC framework} 

\end{figure*}

\section{Proposed Generative Video Semantic Communications Framework}

In this section, we introduce the components of GVSC framework. 
As illustrated in the Fig.~\ref{fig:GVSC framework}, our framework comprises three components: semantic extractor, source channel coding, and GenAI model. 

\subsection{Video Semantic Extractors with Multiple Modalities}
Semantic extraction is essential for video transmission at ultra-low CBR.
Depending on the channel conditions, we propose to extract two kinds of modal semantic information from videos: text description modality and {\color{black}visual} modality.

The text description provides critical movements and events in the video. In our scalable framework, we use a video understanding generative model, such as Video-LLaVA~\cite{DBLP:journals/corr/abs-2311-10122}, to obtain the video description. This model can generate either general or highly detailed video description based on the specific prompt.
While visual semantic information plays a crucial role in video reconstruction by providing detailed structural and spatial context that enhances the consistency of the generated video. {\color{black}Such as the first frame, provides an immediate and comprehensive visual context. This frame offers a rich, color-detailed snapshot that anchors the video's overall aesthetic and thematic elements.
To better save bandwidth, sketch focuses on extracting and conveying the essential contours and major features of each frame. By distilling the video content down to its fundamental visual elements, sketches provide a clear structural guide that aids in reconstructing the basic shapes and layout of scenes. }
This structural framework helps maintain consistency and visual fidelity across frames, ensuring that objects and their spatial relationships are accurately represented throughout the video. 

Denote the input video signal as \(\textbf{X} \in \mathbb{R}^{F \times H \times W \times 3}\), where \(F\) is the number of frames, \(H\) and \(W\) are the height and width of the video, respectively. Let \( \mathbf{x}_f = \textbf{X}_{f,:,:,:} \) represent the \(f\)-th frame of the video for \(1 \leq f \leq F\). 
We can use a sketch extractor to obtain the sketch of \(F\) frames, denoted as \(\textbf{S} \in \mathbb{R}^{F \times H \times W \times 1}\). Let \(\mathbf{s}_f = \textbf{S}_{f,:,:,:} \) represent the \(f\)-th sketch frame for \(1 \leq f \leq F\).

Therefore, even with only a few semantic details as input, the pre-trained GenAI model at the receiver can still reconstruct the video with its main semantic content.

\subsection{Adaptive Transmission Strategy for Multimodal Semantics}
We develop adaptive strategies for transmitting multimodal semantic information, tailoring coding schemes to fit channel conditions and semantic modalities.

\subsubsection{Separate Source Channel Coding for {\color{black}Text Modality}}
Given the importance of video description, textual errors can greatly affect GenAI model's human perception. To minimize the transmission errors of video description semantics, we utilize the separate source channel coding for description, with the code rate $R_c$, based on the different channel conditions.
\subsubsection{Joint Source Channel Coding for {\color{black}Visual Modality}}
Compared with text description modality, {\color{black}visual} modality can better assist GenAI in reconstructing videos.
Moreover, we propose that if the SNR is greater than a predefined threshold, GVSC transmits both description and {\color{black}visual} simultaneously to achieve better reconstruction results. 

To prevent the cliff effect during transmission, we use DJSCC with attention mechanism for sketch sequence transmission. The attention mechanism improves compression efficiency by enabling the network to focus on key semantic information that enhances the overall quality of sketch transmission. Moreover, we design a weighted loss function $\mathcal{L}$ for sketch transmission that combines MSE and LPIPS losses:

\begin{equation} \label{loss func}
\mathcal{L}(x, x_0) = k \cdot \text{MSE}(x, x_0) + (1 - k) \cdot \text{LPIPS}(x, x_0),
\end{equation}
where \( k\) is the weighting factor that balances the importance of the two losses, with the range between [0,1]. By incorporating LPIPS loss, we aim to better optimize the semantic and perceptual quality of the sketches during transmission.
While the widely used MSE metric measures pixel-wise differences, it fails to capture the perceptual significance of the condensed white contours crucial in sketches, necessitating the inclusion of LPIPS to address this shortcoming.

\subsection{Video GenAI Model}

In this section, we delve into the role of GenAI in our framework. We explore the criteria for selecting suitable conditional generation models that enable us to achieve high-quality semantic transmission.

\subsubsection{Latent Space Embeddings}
{\color{black}After passing through the wireless channels, the receiver captures transmitted semantic information, such as video frames, sketches, and descriptions. Our framework processes this diverse content using a pre-trained text encoder (e.g., OpenCLIP ViT-H/14~\cite{cherti2023reproducible}) for textual data and a VAE for visual data like sketches and RGB frames, preparing it for high-quality video reconstruction.}

\subsubsection{Diffusion Process}\label{sec:video diffusion model}
After obtaining the semantic embedding, the video generator utilizes these information as conditions to reconstruct the video at the receiver. During the generation process, the diffusion model \(\epsilon_\theta\) combines the embeddings of the visual semantics and the video description semantics to perform denoising. Assuming \(z_t\) is the latent representation at time step \(t\), the generation process is as follows:
\begin{equation}
  \epsilon_\theta(z_t, c, t) = \epsilon_\theta(z_t, c_{\text{v}}, t) + \omega \left( \epsilon_\theta(z_t, c_{\text{t}}, t) - \epsilon_\theta(z_t, c_{\text{v}}, t) \right),  
\end{equation}
where \(c_{\text{v}}\) and \(c_{\text{t}}\) are the conditional embeddings of the visual information and the text information, respectively, and \(\omega\) is the guidance scale. If the model only has the video description as the conditional input, the process is simplified to
\(\epsilon_\theta(z_t, c_{\text{text}}, t)\).

To ensure consistent output, diffusion models are initialized with the same random seed and hyper-parameters, producing identical content from the same initial noise input when using a deterministic solver. This uniformity is crucial for comparative analysis and applications requiring predictable results.

\begin{figure*}[h]

\centering 
\includegraphics[trim= 200 320 200 30, scale=0.78]{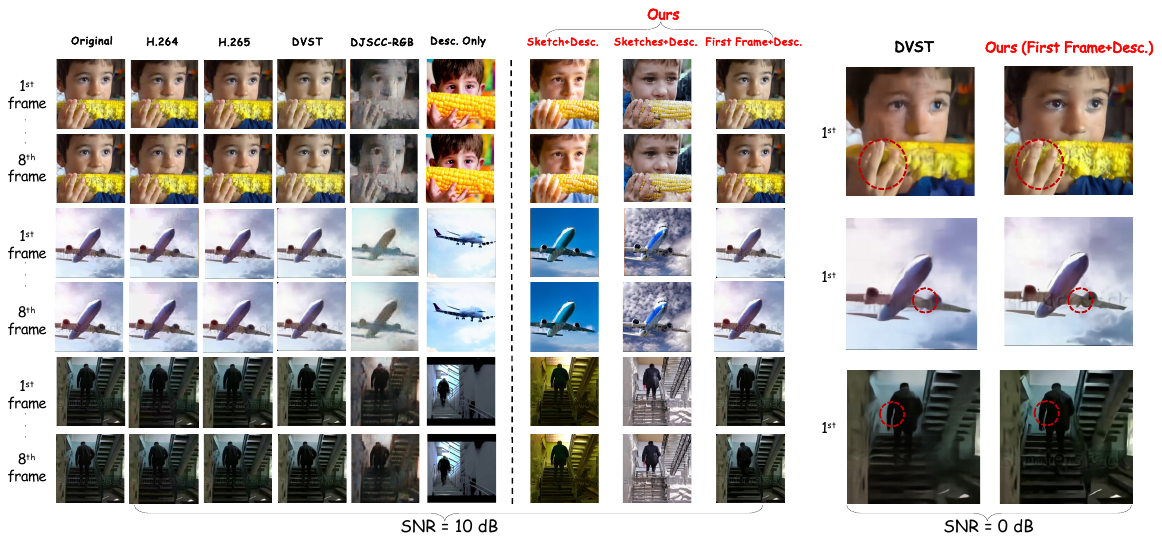}

\caption{{\color{black}Visual comparisons of different transmission schemes at SNR = 10\,dB and SNR = 0\,dB. The first column on the left displays the original video frames. Subsequent columns illustrate the visual outcomes of various schemes at SNR = 10\,dB. The rightmost part focuses on the \textbf{DVST} and \textbf{First Frame+Desc.} schemes specifically at SNR = 0\,dB, all while maintaining a CBR of $10^{-2}$. In order to clearly demonstrate the effect of the best solution and provide a fair performance comparison at 0\,dB, we only selected the best results among the comparison solutions and our proposed solutions for display.}}
\label{fig:case study} 

\end{figure*}

\section{Implementation of GVSC Framework}

In this section, we introduce the specific implementation of the proposed GVSC framework, including transmission strategy, model configuration, and evaluation metrics. 

\subsection{Transmission Strategy and Model Configuration} \label{subsec: cal}
{\color{black}We calculate the widely-adopted {\it channel bandwidth ratio} as $\text{CBR} = \frac{K}{C_{\text{in}}HWF} \quad \text{symbols/pixel}$ \cite{DBLP:journals/jsac/WangDLNSDQZ23, DBLP:journals/jsac/TungG22}.
Here, $K$ denotes the total number of symbols transmitted over the wireless channels, while $C_{\text{in}}$, $H$, $W$, and $F$ correspond to the number of input channels, height, width, and video frames, respectively.}

Considering a low CBR scenario, we set the CBR of the scheme to approximately be the level of $10^{-2}$.
{\color{black}In our research, we develop three semantic transmission strategies, each tailored to leverage different aspects of video content for reconstruction. }

    {\color{black}\textbf{Sketches+Desc.:} In this scheme, we transmit the sketch for each frame to precisely control the layout of objects. Key sketches are transmitted using DJSCC~\cite{dai2022nonlinear}, while the other sketches are sent via deep video semantic transmission (DVST)~\cite{DBLP:journals/jsac/WangDLNSDQZ23}. The average CBR for sketch sequence transmission is 0.0026.} For the transmission of descriptions, we employ turbo coding with a code rate $ R_c = \frac{1}{3} $ and complement it with 4-QAM constellation modulation. Consequently, the number of description symbols to be transmitted is given by $K_d = \frac{N_d}{M R_c} = 1.5 N_d $, where $ N_d $ represents the number of bits for the textual description, $ M $ is the modulation order (which is 2 for 4-QAM), and $ R_c $ is the code rate. Given that the average number of description tokens per video is 95.63, each token typically requires 8 bits, leading to $N_d = 8 \times 95.63 = 765.04 \, \text{bits} $. Therefore, the average number of symbols $K_d = 1.5 \times 765.04 = 1147.56$. The VideoComposer~\cite{DBLP:conf/nips/WangYZCWZSZZ23} subsequently fuses these sketches with their associated descriptions to reconstruct the video, ensuring high fidelity by leveraging extensive semantic information. {\color{black}This scheme achieves an total average CBR of $0.003$.}

    {\color{black} \textbf{Sketch+Desc.:} To enhance the clarity of visual semantics during transmission, we opt for transmitting only the initial frame's sketch, thereby concentrating resources on its quality. An initial sketch of the video is transmitted with video description. The sketch is encoded using the DJSCC optimized for sketch transmission. Stable Diffusion 3.5~\cite{esser2024scaling} is used to reconstruct the first frame of the video, after which Open-Sora~\cite{open_sora_2024} reconstruct the whole video.} We configure the DJSCC-sketch and train the model over an SNR range from 0 to 10\,dB. Testing is performed across all the SNRs within this range. The total number of symbols required for this strategy is given by $K = 1.5N_d + N_s $ where $N_s$ represents the number of transmitted symbols for the sketch, and $1.5N_d = 1147.56$ symbols are required for the video description. The output size of the DJSCC encoder is $32 \times 32 \times 2$. Therefore, the resource blocks allocated for the sketch data are calculated as $\frac{32 \times 32 \times 2}{2} = 1024 $ i.e., resource blocks for sketch. Thus, the total number of resource blocks required for the \textbf{Sketch+Desc.} scheme is:$ K = 1147.56 + 1024 = 2171.56 $. Consequently, the CBR for this scheme is $0.001$.

    {\color{black}\textbf{First Frame+Desc.:} For acquiring more detailed semantic information, we choose to transmit the first frame with video descriptions, which are transmitted using DJSCC. Open-Sora then utilizes these rich semantic inputs to reconstruct the video. This scheme combines a high-quality visual anchor with textual semantics to detail dynamic content, ensuring the high-quality video reconstruction. For this scheme, we have specifically trained the DJSCC to optimize the transmission of the first frame. The video description is transmitted as \textbf{Desc. Only} scheme. This scheme achieve a average CBR of $0.0031$.}

\subsection{Evaluation Metrics}

The GVSC Framework adopts the following metrics to evaluate semantic similarity.
\textbf{CLIP score:} {\color{black} The CLIP score~\cite{radford2021learning}, widely adopted due to its training on large image-text datasets \cite{careil2023towards, lei2023text+, fan2024semantic, li2024misc}, effectively captures high-level semantic features and is suitable for assessing semantic similarity.} We use the average CLIP score to compare each generated frame with its corresponding original frame. It effectively measures the semantic similarity of each video frame. The CLIP score focuses on frame-level semantic similarity, ensuring each generated frame matches its original counterpart. \textbf{BERT score:} We design a video semantic consistency check scheme. To assess overall semantic similarity, we use Video-LLaVA to generate captions for both videos and evaluate their similarities with a pretrained BERT model~\cite{zhang2019bertscore}. Given Video-LLaVA's generative diversity, we sample captions 3 times for each video and average the scores for the overall semantic score. The BERT score evaluates the video-level semantic consistency by comparing generated captions, measuring how well the generated video captures the broader narrative and context.

\textbf{PSNR and SSIM~\cite{wang2004image}:} The peak signal-to-noise ratio (PSNR) and structure similarity index measure (SSIM) are traditional video quality metrics, providing insights into pixel and structure-level fidelity. They measure the peak SNR and structural similarity, respectively.

\section{Simulations}

In this section, we introduce the datasets used, the comparison models, the evaluation schemes and the experimental results in detail.

\subsection{Simulation Setups}
\subsubsection{Dataset}
We use the WebVid dataset~\cite{DBLP:conf/iccv/BainNVZ21} for both training and testing. WebVid contains Internet-crawled videos with captions covering diverse scenes such as human actions and natural landscapes. We use the captions as semantic descriptions, with each video containing $F=8$ frames at a resolution of $256 \times 256$ ($H = W = 256$). For training the transmission model, we extract over 11,000 sketches using PiDiNet~\cite{DBLP:conf/iccv/0002LYH00P021}, and 129 videos for testing.

\subsubsection{Simulation Parameters}
For transmitting sketches, we train the DJSCC for 200 epochs using the Adam optimizer and batchsize is 16. The learning rate is $1 \times 10^{-4}$. For transmitting first frame, we adjust the learning rate to $1 \times 10^{-5}$.
Following the human assessment simulations, we set \( k = 0.3 \) in~(\ref{loss func}). 
We train our transmission model over a SNR range of 0 to 10\,dB and test all models within this range. Due to the excessive number of parameters, we freeze the GenAI model. We set the frame rate to 4, the inference time-step to 50, and the random seed to 7777.

We use an additive white Gaussian noise (AWGN) channel, where the noise vector \(\textbf{n}\) is modeled as a complex Gaussian distribution \(\textbf{n} \sim \mathcal{CN}(0, \sigma^2 \textbf{I})\), with zero mean and covariance \(\sigma^2 \textbf{I}\). Here, \(\sigma\) represents the standard deviation of the noise, and \(\textbf{I}\) is the identity matrix whose dimensions correspond to the number of dimensions in \(\textbf{n}\). 

\subsection{Comparison Schemes}

{\textbf{H.264/H.265+LDPC:} We use H.264 and H.265 for source coding, with LDPC codes and QAM modulation for channel coding. LDPC rates and modulation orders are adapted to the SNR, while keeping CBR = 0.005 to ensure fair comparison. Detailed configurations are listed in Table~\ref{tab:ldpc_config}.}

\begin{table}[htbp]

\centering
\caption{LDPC and modulation configurations under different SNRs}

\resizebox{\linewidth}{!}{
\begin{tabular}{c|ccc}
\toprule
\textbf{CBR} & \textbf{0 dB} & \textbf{2 dB} & \textbf{4 dB} \\
\midrule
0.005 & 1/3 LDPC+4QAM & 1/2 LDPC+4QAM & 2/3 LDPC+4QAM \\
\bottomrule
\end{tabular}
}

\resizebox{\linewidth}{!}{
\begin{tabular}{c|ccc}
\toprule
\textbf{CBR} & \textbf{6 dB} & \textbf{8 dB} & \textbf{10 dB} \\
\midrule
0.005 & 3/4 LDPC+4QAM & 1/2 LDPC+16QAM & 2/3 LDPC+16QAM \\
\bottomrule
\end{tabular}
}

\label{tab:ldpc_config}
\end{table}

\color{black}
\textbf{DJSCC-RGB:} 
We use DJSCC with three layers of upsampling and downsampling modules. We maintain the original channel setting (3 channels) to transmit the RGB frames of the videos. The average CBR of this scheme is $0.005$.

\textbf{Desc. Only:}
We employ turbo coding with a coding rate $ R_c = \frac{1}{3} $ and 4-QAM constellation modulation to transmit video descriptions. At the receiver, Open-Sora~\cite{open_sora_2024} reconstructs the video solely from these descriptions, demonstrating the effectiveness of text-based video generation. The average CBR of this scheme is $0.0007$.

\textbf{DVST:}
To leverage the inter-frame dependencies effectively, we utilizes nonlinear transform source channel coding (NTSCC)~\cite{dai2022nonlinear} for key frames, while \textbf{DVST}~\cite{DBLP:journals/jsac/WangDLNSDQZ23} handles the other frames. This scheme maintains an average CBR of $0.004$.

\color{black}
\begin{figure}[t]
    \centering

    \includegraphics[trim= 25 80 100 50, clip, scale=0.34]{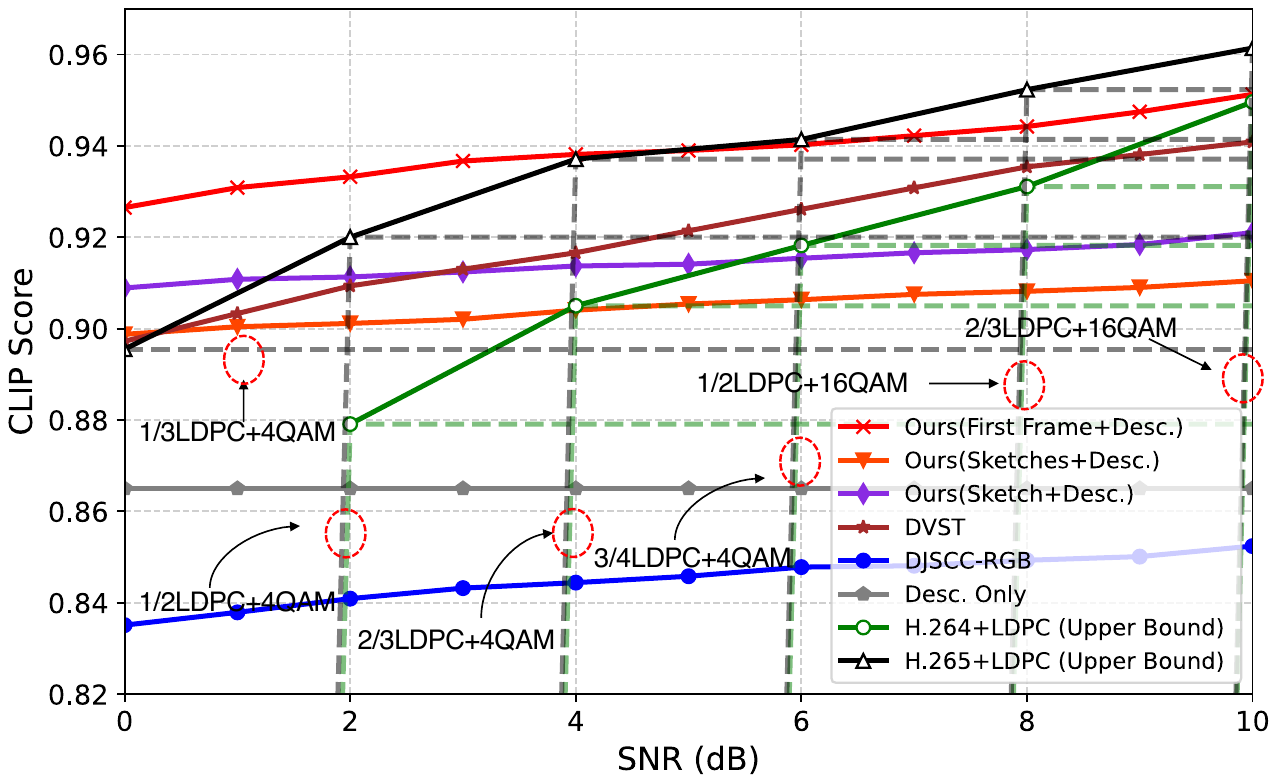}

    \caption{{CLIP score of different schemes for various SNRs.}}
    \label{fig:snr-CLIP}

    \includegraphics[trim= 25 80 100 50, clip, scale=0.34]{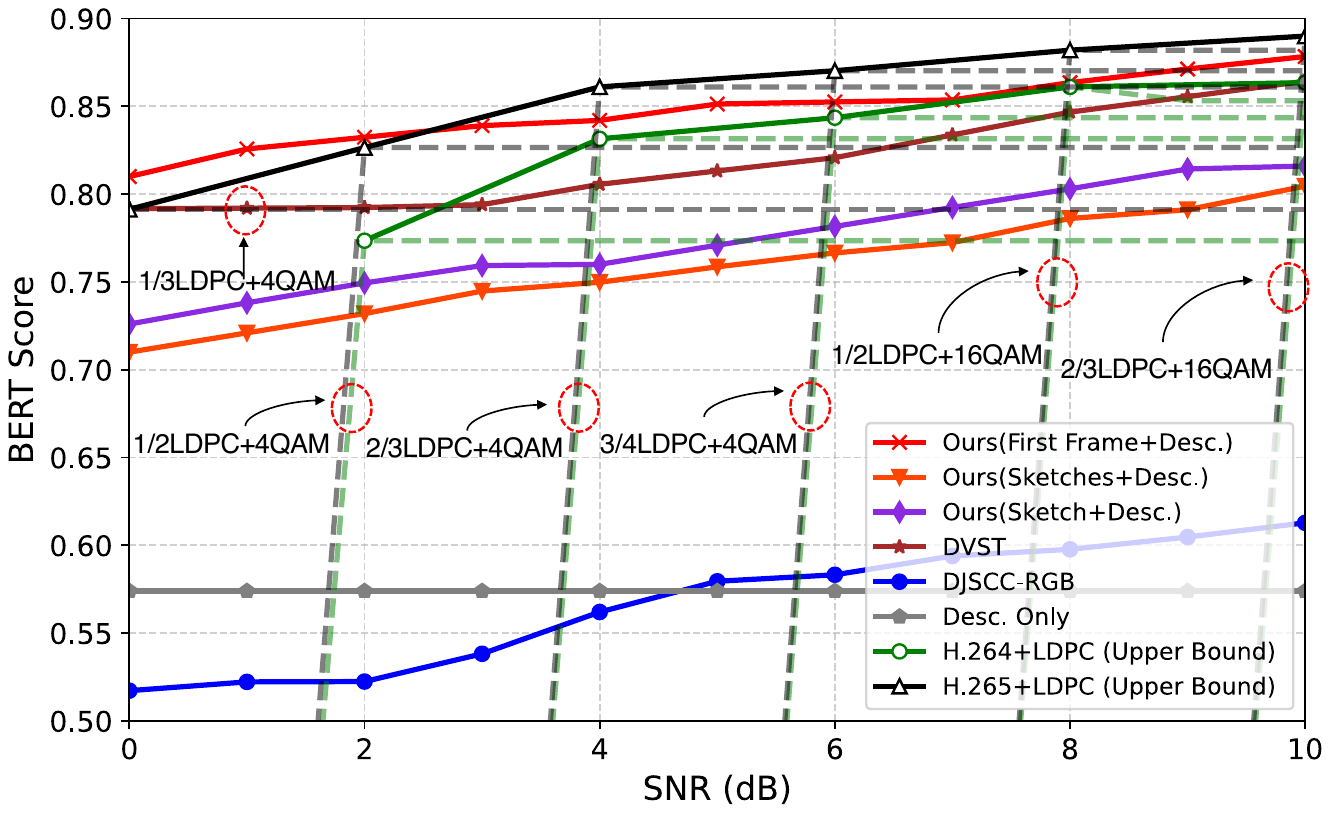}

    \caption{{BERT score of different schemes for various SNRs.}}
    \label{fig:snr-BERT}

\end{figure}

\begin{figure}[t]

    \centering
    \includegraphics[trim= 25 80 100 50, clip, scale=0.34]{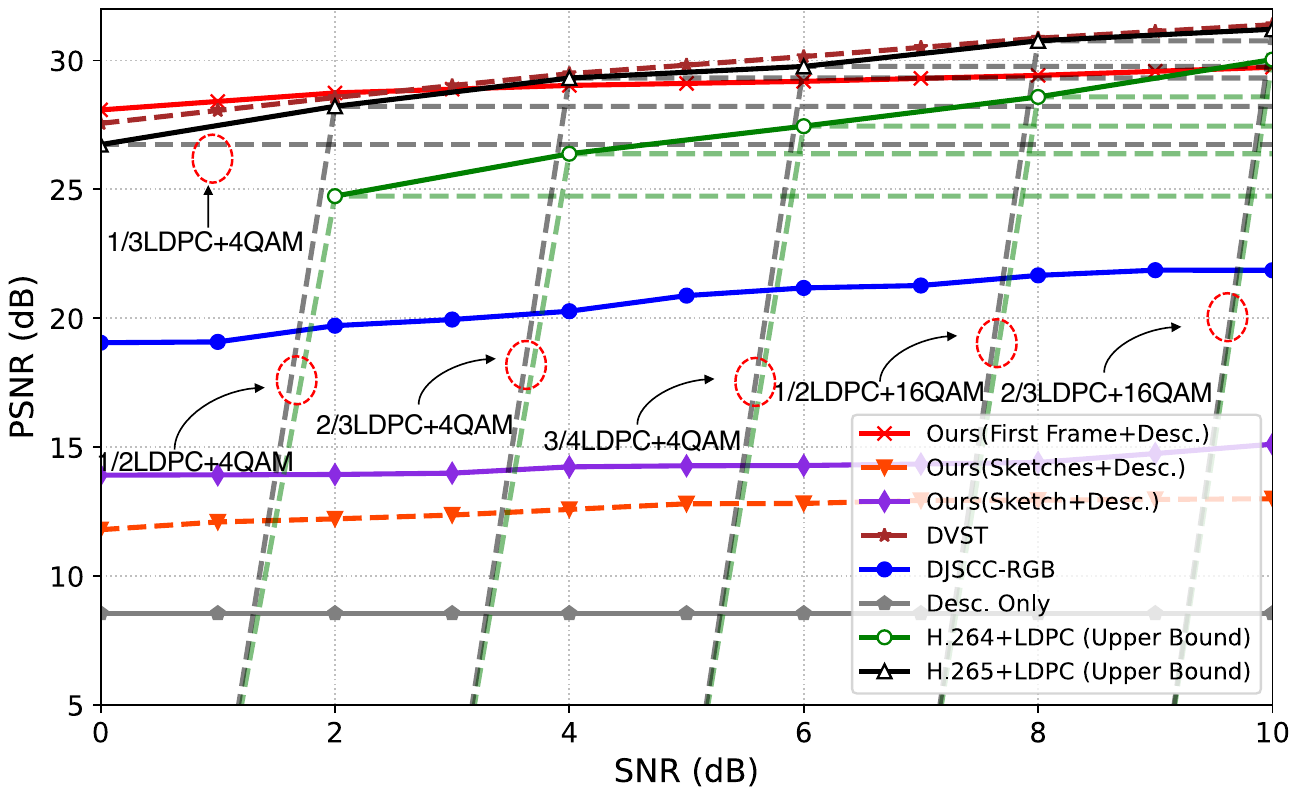}

    \caption{{PSNR of different schemes for various SNRs.}}
    \label{fig:snr-psnr}

\end{figure}

\begin{figure}[t]

    \centering
    \includegraphics[trim= 25 80 100 50, clip, scale=0.34]{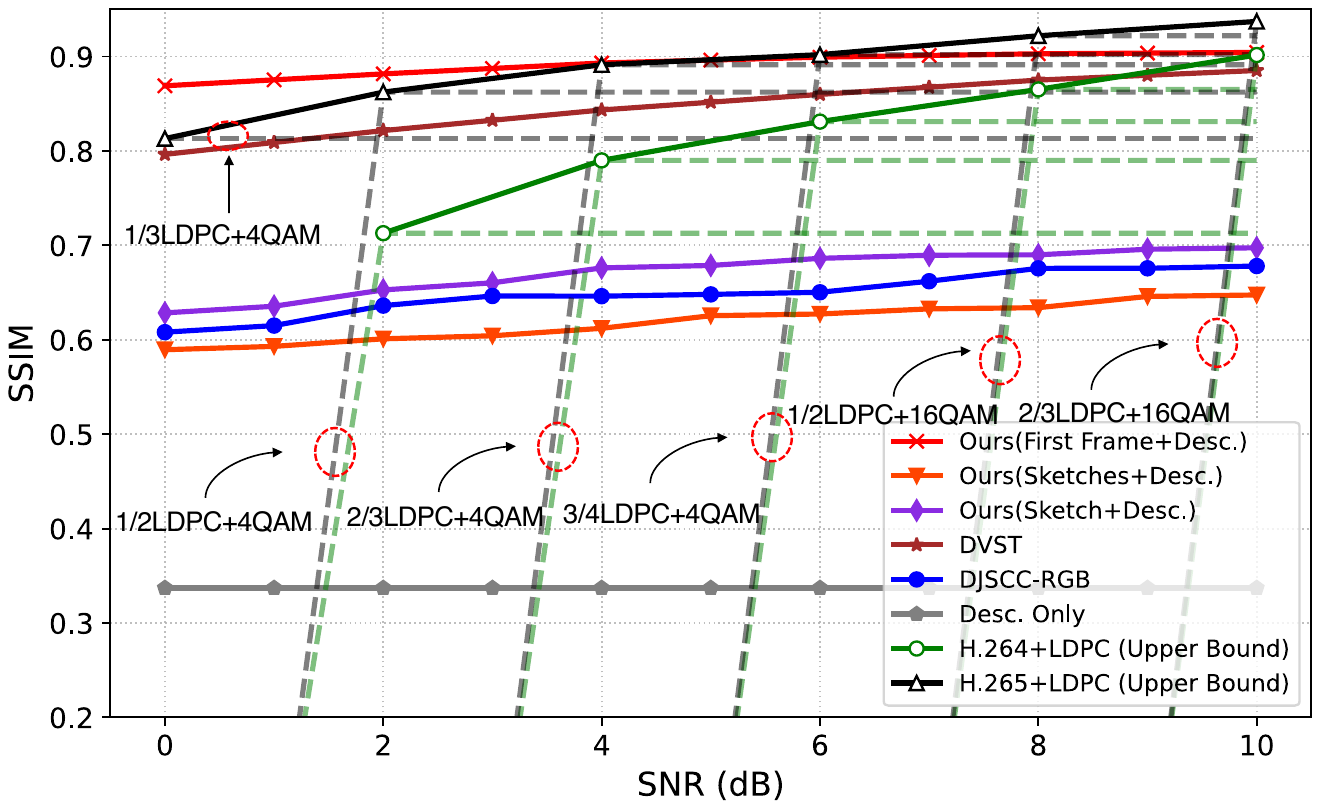}

    \caption{{SSIM of different schemes for various SNRs.}}
    \label{fig:snr-SSIM}

\end{figure}

\subsection{Simulation Results}

\color{black}{To evaluate the performance of our proposed video semantic extraction strategies, we conducted extensive simulations across a range of SNR conditions. Fig.~\ref{fig:case study} shows the visualization of different schemes. Fig.~\ref{fig:snr-CLIP}, Fig.~\ref{fig:snr-BERT}, Fig.~\ref{fig:snr-psnr} and Fig~\ref{fig:snr-SSIM} present the comparison results in terms of CLIP score, BERT score, PSNR, and SSIM, respectively.}

Visual comparisons of different transmission schemes at SNR = 10\,dB and 0\,dB are shown in Fig.~\ref{fig:case study}. The first column on the left displays the original video frames (the 1st and the 8th frames). Subsequent columns present visual results of various methods at SNR = 10\,dB. The right part highlights the \textbf{DVST} and \textbf{First Frame+Desc.} schemes at SNR = 0\,dB, both maintaining a CBR of $10^{-2}$. \textbf{H.264+LDPC} and \textbf{H.265+LDPC} are able to retain structural details in high SNR. DVST shows the effective transmission at SNR = 10\,dB with high visual quality, while \textbf{DJSCC-RGB} undergoes significant visual degradation due to ultra-low CBR, showing highly distorted frames. The \textbf{Desc. Only} scheme preserves basic semantic information but loses structural details, leading to relatively poor video reconstruction. Conversely, the \textbf{Sketch+Desc.} and \textbf{Sketches+Desc.} schemes excel in retaining both structural and textual semantics at lower CBR, as reflected in descriptions such as \textit{``A boy is eating corn"}. The \textbf{First Frame+Desc.} scheme allocates more resources to the first frame, ensuring semantic consistency and excelling in visual clarity even at SNR = 0\,dB, as highlighted in the red circles. 

\begin{figure}[t]
    \centering
    \includegraphics[trim= 0 0 40 30, clip, scale=0.3]{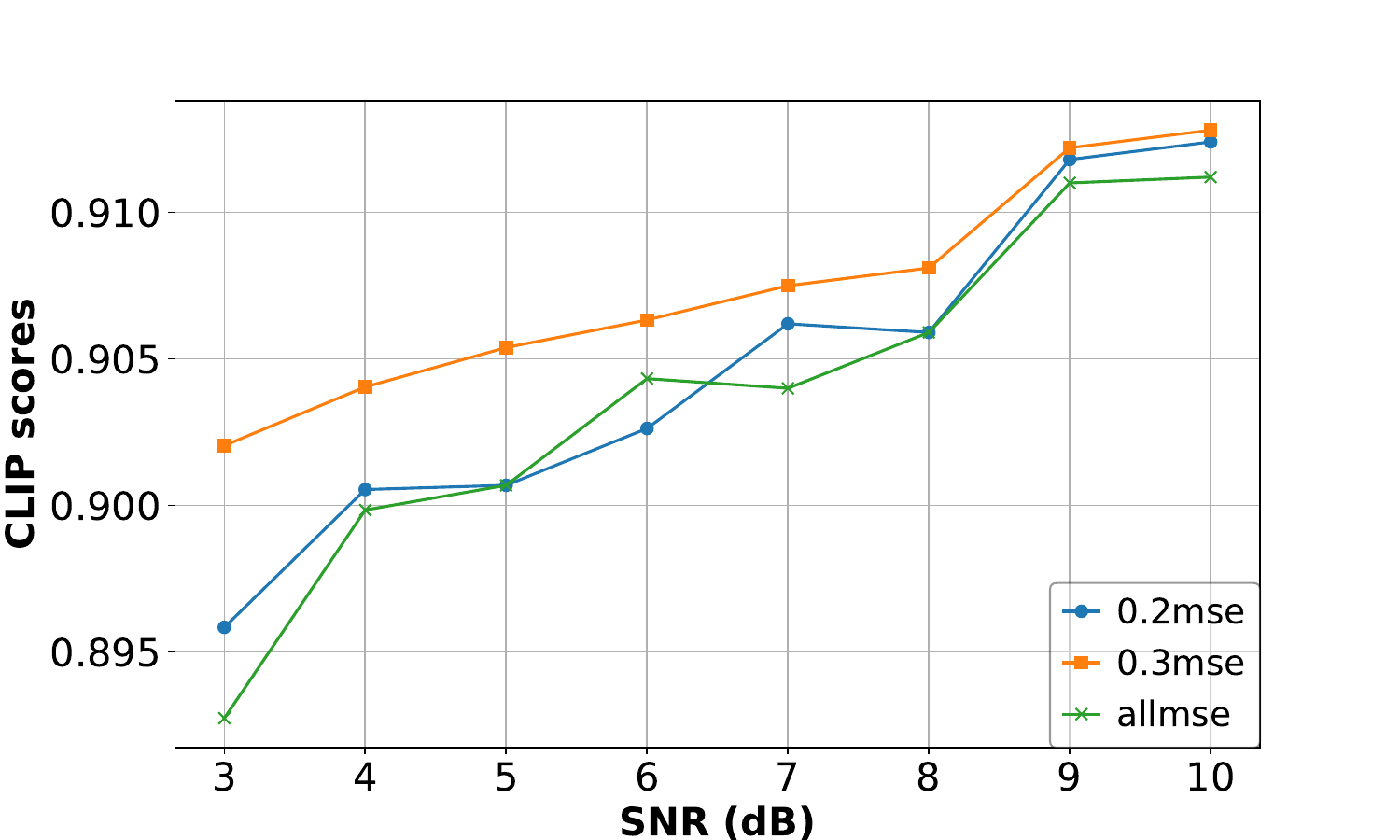}

    \caption{CLIP score for various SNR\textcolor{black}{s} with different \textit{k} in \textbf{Sketch+Desc.} scheme.}
    \label{fig:mselpips}

\end{figure}

{In Figs.~\ref{fig:snr-CLIP},~\ref{fig:snr-BERT},~\ref{fig:snr-psnr} and~\ref{fig:snr-SSIM}, the red circles in these figures correspond to different LDPC code rates and modulation schemes used in our experiments.
\color{black}
Fig.~\ref{fig:snr-CLIP} shows the CLIP scores of different schemes across SNR levels.
With adaptive coding, \textbf{H.264+LDPC} fails to work at 0\,dB, and \textbf{H.265+LDPC} still underperforms compared to \textbf{First Frame+Desc.} under low SNR conditions. 
\color{black}
\textbf{DJSCC-RGB} and \textbf{Desc. Only} are constrained by the lack of visual context, while \textbf{Sketch+Desc.} and \textbf{Sketches+Desc.} benefit from structural information and show improvements. \textbf{DVST} improves gradually with increasing SNR, but is eventually surpassed by \textbf{H.265+LDPC} at higher SNRs. Notably, \textbf{First Frame+Desc.} consistently achieves the highest performance in all SNRs, demonstrating the robustness of GVSC framework under challenging channel conditions.

Fig.~\ref{fig:snr-BERT} presents BERT score performance of different schemes for various SNR conditions. {Similar to CLIP score, \textbf{H.264+LDPC} has difficulty in reliably conveying the at SNR $=$ 0\,dB.} The BERT scores for \textbf{Sketch+Desc.} and \textbf{Sketches+Desc.} schemes are lower than \textbf{H.264+LDPC} and \textbf{H.265+LDPC} because traditional schemes transmit detailed visual information closely related to the original video, leading to higher text similarity scores. When generating descriptions, the visual model might include non-critical details such as hair color or eye direction, which are not always relevant in the video generation context. Since BERT focuses on specific textual details, even slight discrepancies can result in lower scores. We can conclude that the semantic information of different modalities has a significant impact on the BERT score. Specifically, the \textbf{Desc. Only} scheme has the least semantic information, and \textbf{DJSCC-RGB} is heavily affected by low SNR. \textbf{Sketch+Desc.} and \textbf{Sketches+Desc.} schemes maintain basic control over structural and action semantics, so the BERT score is relatively high. {When SNR exceeds 2\,dB, \textbf{H.264+LDPC}, \textbf{H.265+LDPC}, and \textbf{First Frame+Desc.} achieve similar BERT scores, indicating that these scheme provide comparable support for coarse-grained caption generation under moderate-to-good channel conditions.}

Fig.~\ref{fig:snr-psnr} presents the PSNR performance of different schemes for various SNRs. Compared with other performance metrics, PSNR focuses primarily on pixel-level reconstruction quality. As depicted in Fig.~5, methods lacking strong semantic control capabilities usually struggle to achieve pixel level consistency, resulting in the lower PSNR scores. Although \textbf{DJSCC-RGB} may appear blurry, it achieves higher pixel similarity than \textbf{Sketch+Desc.} and \textbf{Sketches+Desc.} schemes. However, schemes using the first frame as semantic information still perform effectively on the PSNR, demonstrating the robust pixel-level accuracy.

As illustrated in Fig.~6, SSIM assesses the structural differences in video content. The \textbf{Desc. Only} method lacks structured semantic controls and has the lowest SSIM score as shown in Fig. 6, indicating its poor performance in structural integrity. In contrast, methods utilizing sketches exhibit a gradual increase in SSIM scores with improving SNR, reflecting the enhanced ability to convey structural semantic information. This upward trend in SSIM scores highlights the capability of \textbf{Sketch+Desc.} and \textbf{Sketches+Desc.} schemes to improve structural fidelity in video reconstruction, thus enhancing the overall quality and viewing experience. Moreover, the \textbf{First Frame+Desc.} scheme, which utilizes the most comprehensive semantic information, exhibits the strongest performance in terms of structural semantics. This is reflected in its consistently high SSIM scores in all SNR levels, highlighting its effectiveness in preserving both the integrity and detail of video structure, thereby providing a compelling case for its use in scenarios that demand high fidelity in semantic and structural aspects.

Fig.~\ref{fig:mselpips} shows the CLIP score performance across various SNR conditions for different sketch transmission loss functions. As SNR increases, CLIP score improve for all configurations, indicating better semantic consistency and visual fidelity. The configuration with \(k = 0.3\) consistently achieves the highest CLIP score, suggesting that balancing MSE and LPIPS loss leads to optimal performance.

\section{Conclusion}

In this work, we proposed the GVSC, a novel framework to video semantic communication using GenAI large models. It integrates video semantic extraction, source-channel transmission, and semantic encoding/decoding/generation. Video semantics is still an emerging concept, with varying interpretations across different tasks. We define video semantics as consistent semantic information during transmission. To handle diverse transmission needs, we designed multimodal strategies. Our method reduces transmission resources and achieves high CLIP, BERT, PSNR, and SSIM scores. We also design a weighted loss to improve sketch transmission. Extensive simulations show that GVSC works well at low CBR, demonstrating the feasibility of GenAI for efficient bandwidth-limited video transmission. Although it has higher latency than traditional methods, it is suitable for low-bandwidth and delay-tolerant scenarios. With future model optimizations, GVSC may support real-time applications and benefit future communication systems.

\bibliography{reference}
\bibliographystyle{IEEEtran}

\end{document}